\documentclass[journal=jpclcd, layout=twocolumn]{achemso}

\usepackage[T1]{fontenc} 

\usepackage{amsmath,color, tikz}
\usetikzlibrary{matrix}
\usepackage{amssymb}
\usepackage{amsfonts}
\usepackage{amsthm}
\usepackage{mathtools}
\usepackage{comment}

\usepackage{xspace}

\usepackage{algorithm}
\usepackage{algpseudocode}
\usepackage{dsfont}

\usepackage{bbold}
\usepackage{chemformula}
\usepackage{siunitx}

\DeclarePairedDelimiterX\braket[2]{\langle}{\rangle}{#1 \delimsize\vert #2}
\DeclarePairedDelimiterX\braket3[3]{\langle}{\rangle}{#1 \delimsize\vert #2 \delimsize\vert #3}

\usepackage{accents}

\usepackage{fancyvrb}

\newcommand{\MaxwellLink}{\textsc{MaxwellLink}\xspace}

\usepackage{pdfpages}
\makeatletter
\AtBeginDocument{\let\LS@rot\@undefined}
\makeatother

\usepackage{hyperref}
\hypersetup{
	colorlinks=true,
	citecolor=blue,
	linkcolor=red,
	urlcolor = blue
}
\usepackage{cleveref}

	\title{Atomistic Emergence of Vibrational Polariton Condensation}

    \author{Xinwei Ji}
    \affiliation{Department of Physics and Astronomy, University of Delaware, Newark, Delaware 19716, USA}
	
	\author{Tao E. Li}%
	\email{taoeli@udel.edu}
	\affiliation{Department of Physics and Astronomy, University of Delaware, Newark, Delaware 19716, USA}
	
\begin{document}

\begin{tocentry}
    \includegraphics[width=1.0\linewidth]{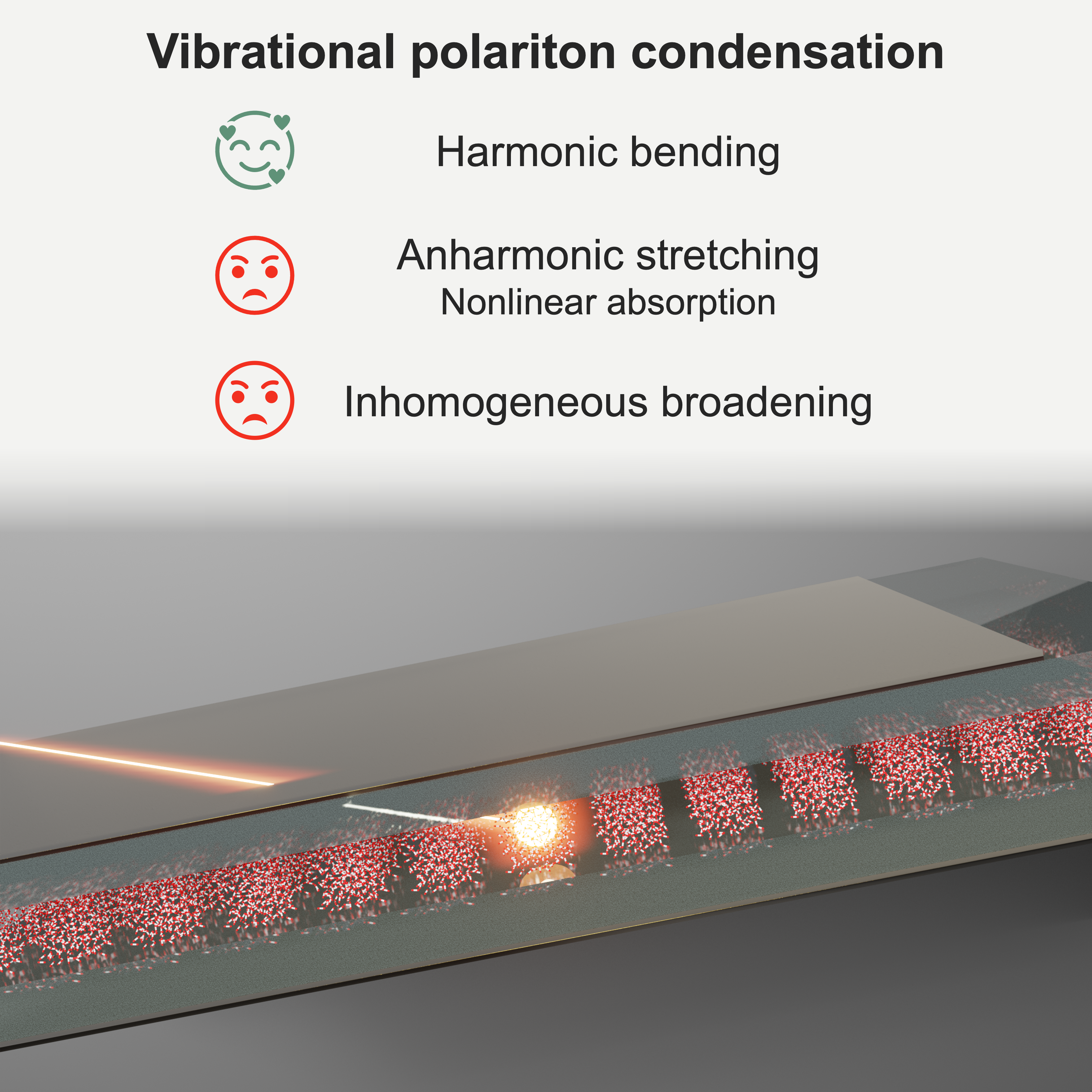}
\end{tocentry}

    \begin{abstract}
Nonequilibrium polariton condensation is a macroscopic quantum phenomenon in which overwhelming population accumulation is observed at the  band edge of the lower polariton (LP) dispersion curve under incoherent material pumping. As a promising route toward nonequilibrium Bose--Einstein condensates, its description, however, remains predominantly limited to models with  phenomenological material parameters.   Herein, a fully atomistic description of vibrational polariton condensation is developed  by employing mesoscale cavity molecular dynamics simulations, a method which self-consistently propagates classical electromagnetic dynamics coupled to realistic molecules in a multimode Fabry--P\'erot cavity.  Under continuous-wave pumping, a  lower polariton condensate  forms when the infrared cavity is coupled to harmonic, homogeneously broadened molecular vibrations, such as the bending modes of liquid \ch{CO2} and \ch{CS2}, whereas anharmonic stretching modes and liquids with fast intramolecular relaxation hardly condense. The role of libration-bending coupling for facilitating the LP condensation is further discussed.
GPU-accelerated mesoscale  simulations involving approximately 50 million atoms additionally provide direct two-dimensional real-space imaging of fluid-like polariton patterns during the condensation formation. Overall, this study provides key design insights into vibrational polariton condensation as a novel approach for controlling liquid-phase chemistry.
\end{abstract}


\section*{Introduction}

Polaritons are hybrid bosonic quasiparticles originating from the coherent coupling between elementary excitations in matter and cavity photons under strong light-matter interactions  \cite{huangInteractionRadiationField1951,Hopfield1958,hopfieldPolaritonAbsorptionLines1965,Weisbuch1992,Shalabney2015,Long2015,Wright2023,Brawley2025NatChem}. In semiconductor microcavities, the hybridization of excitons and cavity photon modes gives rise to exciton polaritons \cite{Weisbuch1992,Deng2010,Carusotto2013,szymanskaNonequilibriumQuantumCondensation2006}. With exciton-polaritons,  the first demonstration of nonequilibrium Bose--Einstein condensation was experimentally reported  twenty years ago \cite{kasprzakBoseEinsteinCondensation2006,utsunomiyaObservationBogoliubovExcitations2008}. As a driven--dissipative nonequilibrium state, exciton polariton condensation provides a paradigmatic platform for studying nonequilibrium quantum many-body physics \cite{lagoudakisQuantizedVorticesExciton2008,kena-cohenRoomtemperaturePolaritonLasing2010,nitscheAlgebraicOrderBerezinskiiKosterlitzThouless2014,plumhofRoomtemperatureBoseEinstein2014,fontaineKardarParisiZhang2022,ardizzonePolaritonBoseEinstein2022,castellanosNonEquilibriumBoseEinstein2023,widmannObservationKardarParisiZhangUniversal2026}, such as the Bogoliubov excitations and quantized vortices \cite{utsunomiyaObservationBogoliubovExcitations2008,lagoudakisQuantizedVorticesExciton2008}, Berezinskii–Kosterlitz–Thouless transition \cite{nitscheAlgebraicOrderBerezinskiiKosterlitzThouless2014}, and Kardar–Parisi–Zhang  universality \cite{fontaineKardarParisiZhang2022,widmannObservationKardarParisiZhangUniversal2026}. 
Despite this exciting experimental progress, a unified mechanistic understanding of exciton-polariton condensation has not been established. While existing theoretical studies have provided key insights into the underlying mechanisms \cite{szymanskaNonequilibriumQuantumCondensation2006,woutersExcitationsNonequilibriumBoseEinstein2007,woutersSpatialSpectralShape2008,dagvadorjNonequilibriumPhaseTransition2015,altmanTwoDimensionalSuperfluidityExciton2015a,deligiannisKardarParisiZhangUniversalityDiscrete2022,perez-sanchezRadiativePumpingVs2025}, the use of reduced phenomenological models on the matter side, such as  the generalized Gross-Pitaevskii equation \cite{woutersExcitationsNonequilibriumBoseEinstein2007} and  the Holstein-Tavis–Cummings (HTC) Hamiltonian \cite{Keeling2020}, hinders the predictive design of polariton condensation with varied material systems.

During the past decade, molecular vibrational polaritons have emerged as a promising technology for controlling various physical and chemical processes in the electronic ground state \cite{Shalabney2015,Long2015,Wright2023,Brawley2025NatChem}, including thermally-activated chemical or crystallization processes \cite{Thomas2016, Thomas2019_science,Hirai2020Crys,Ahn2023Science}, as well as infrared (IR) photochemistry \cite{Chen2022,Yin2025}. In this vibrational strong coupling (VSC) regime, existing theoretical work investigates possible routes toward macroscopic vibrational polariton condensation as a novel approach for controlling chemistry \cite{pannir-sivajothiDrivingChemicalReactions2022,shishkovSympatheticMechanismVibrational2024,mondalMacroscopicCondensationTheory2026}. In detail, if a macroscopic population could be driven into a single vibrational polariton mode, the resulting condensate would constitute a nonequilibrium, mode-selective reservoir of vibrational energy distribution for enabling new chemistry. Pioneered by Yuen-Zhou and co-workers, the use of vibrational polariton condensation is studied for enabling pumping-dependent electron-transfer reactions \cite{pannir-sivajothiDrivingChemicalReactions2022}. However, as these works rely mostly on model parameters, it remains unclear whether vibrational polariton condensation may exist and what material systems may provide advantageous platforms for vibrational polariton condensates.

\begin{figure}
    \centering
    \includegraphics[width=0.9\linewidth]{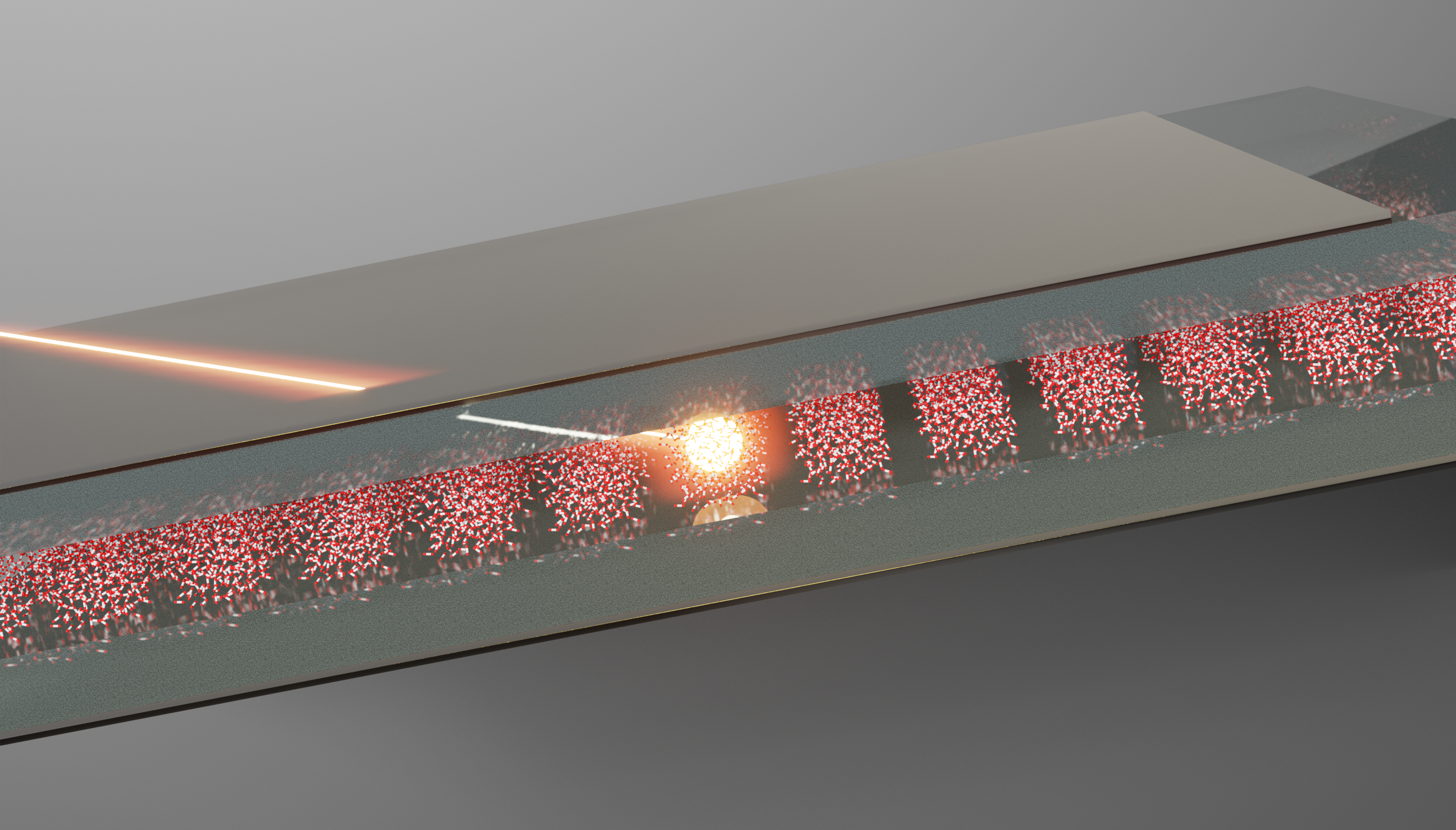}
    \caption{Schematic diagram of mesoscale CavMD simulations of vibrational polariton condensation.}
    \label{fig:pumping_and_cavity}
\end{figure}

Here, we report the atomistic design principles of nonequilibrium vibrational polariton condensation in planar Fabry--P\'{e}rot cavities. Beyond existing studies of model systems, we employ a fully atomistic mesoscale cavity molecular dynamics (CavMD) scheme \cite{Li2020Water,Li2024CavMD,jiNonlinearFreezingVibrational2026a}. In this approach, as depicted in Fig. \ref{fig:pumping_and_cavity}, we self-consistently solve the coupled classical dynamics of up to $\sim$0.5 million independent molecular dynamics cells coupled to a multimode cavity, with each molecular cell being a liquid-phase system and experiencing spatially varied electromagnetic fields. The details of mesoscale CavMD are provided in Supporting Information (SI) Sec. S1, and the associated simulation parameters are listed in SI Sec. S2.

\section*{Results}

We first examine the possibility of nonequilibrium polariton condensation for liquid \ch{CO2} confined in a one-dimensional (1D) Fabry--P\'erot cavity at room temperature. Vibrational polariton dynamics of liquid \ch{CO2} have been extensively studied using CavMD, with key fundamental polariton mechanisms in good agreement with experiments, such as polariton dephasing to molecular dark states \cite{Li2021Relaxation,Xiang2019} and polariton-enhanced molecular nonlinear absorption \cite{Li2020Nonlinear,Xiang2021JCP}. The mesoscale Fabry--P\'erot cavity simulations further reveal ballistic-to-diffusive transport of vibrational polaritons \cite{jiNonlinearFreezingVibrational2026a,Bhakta2026}. For tackling experimental-scale systems, here, the molecules are described by a flexible non-polarizable force field, although more advanced molecular theories, spanning from quantum-mechanical molecular-mechanical (QM/MM) simulations \cite{Li2023QMMM}, density-functional tight-binding \cite{wickramasingheOntheflyCavityMolecular2026,wickramasingheModeSelectiveAnharmonicityDependentEnergy2026}, to machine-learning potentials \cite{liCboamdMachineLearning2026}, have also been used for propagating CavMD. 

\begin{figure}
\centering
\includegraphics[width=\linewidth]{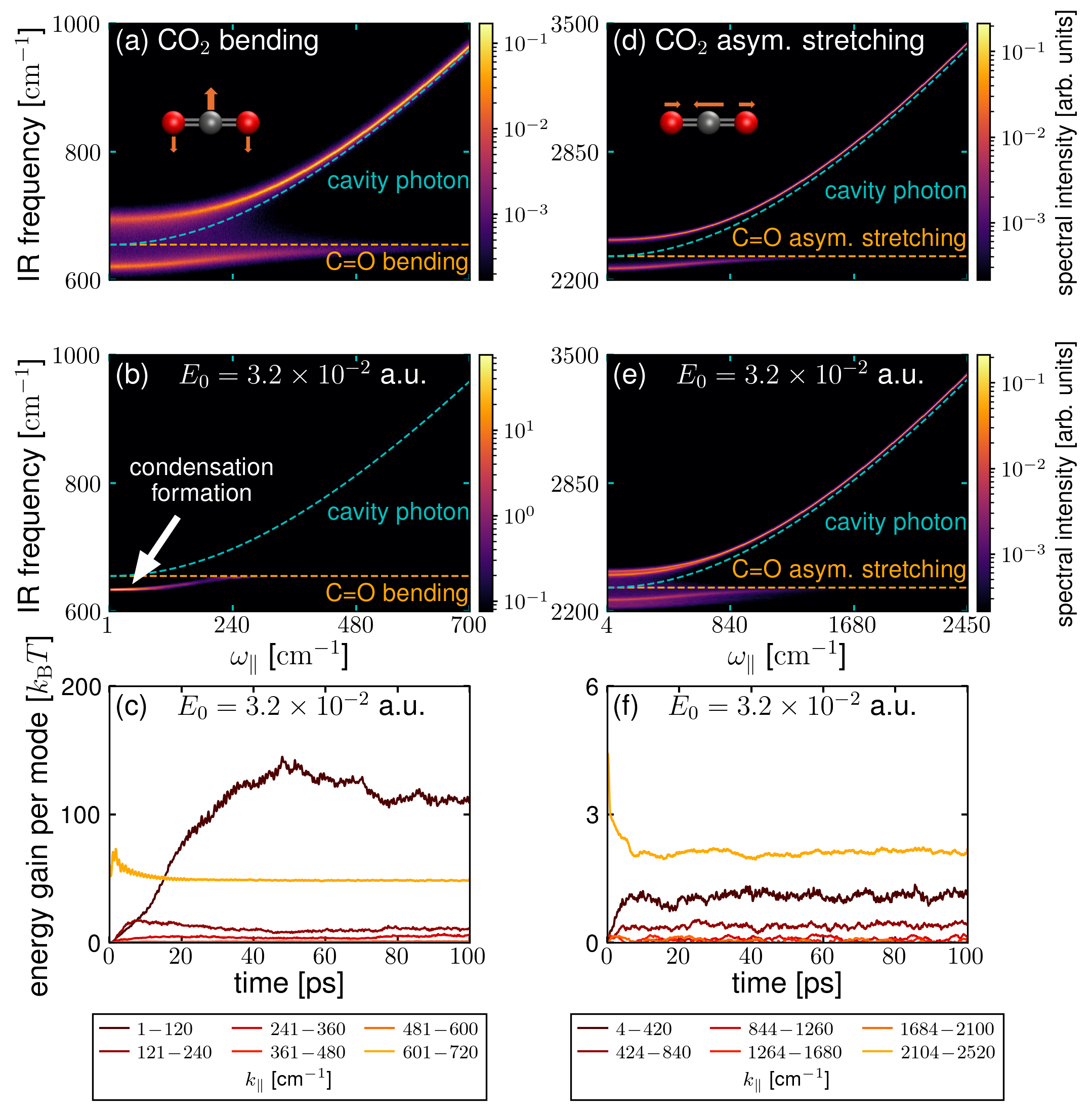}
\caption{(a) Equilibrium polariton dispersion when the \ch{CO_2} bending mode forms strong coupling. (b) Corresponding nonequilibrium polariton dispersion under strong cw pumping ($E_0=3.2\times 10^{-2}$ a.u.). (c) Associated photonic energy dynamics under strong pumping. (d-f) The analogous (d) equilibrium polariton dispersion, as well as (e) nonequilibrium dispersion and (f) photonic energy dynamics under strong pumping for the asymmetric stretching mode of \ch{CO2} under strong coupling. LP condensation forms only for the bending-mode polaritons.}
\label{fig:1}
\end{figure}

\begin{figure*}[h]
\centering
\includegraphics[width=1\linewidth]{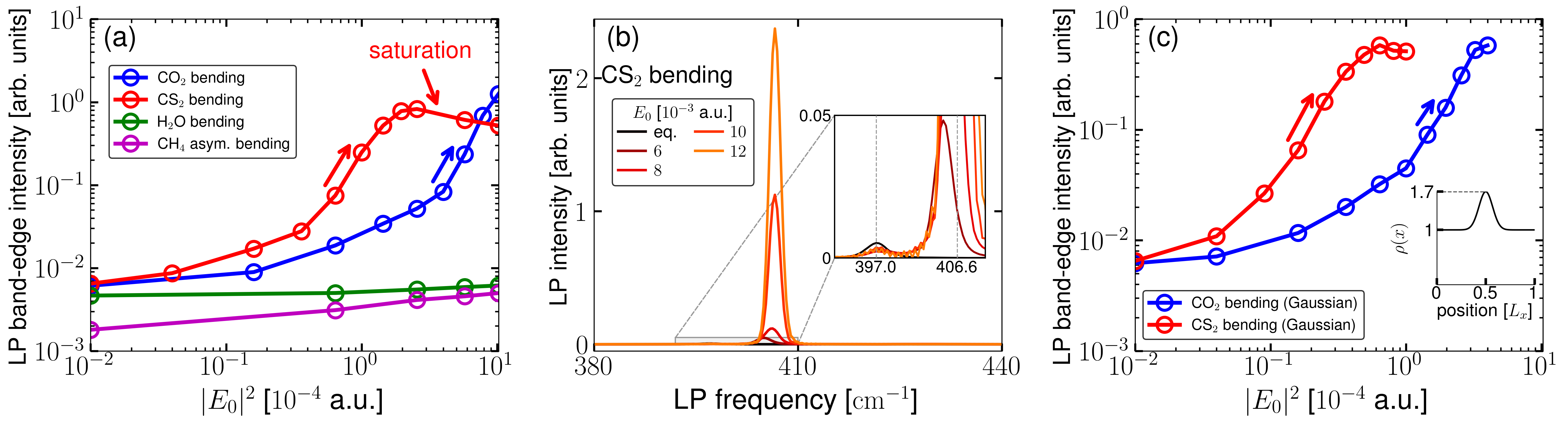}
\caption{Band-edge LP intensity versus the pumping power $|E_0|^2$ when the bending mode of molecules forms polaritons. (a) Four different molecular liquids are compared: \ch{CO2} (blue), \ch{CS2} (red), \ch{H2O} (green), and \ch{CH4} (magenta). The arrows highlight the superlinear trend toward the LP condensation. (b) Corresponding $k_{\parallel}\sim 0$ polariton spectra for the \ch{CS_2} bending mode forming strong coupling under different pumping amplitudes. (c) Analogous calculations as part (a) for \ch{CO2} and \ch{CS2} when the molecular density distribution follows a non-uniform Gaussian distribution along the cavity mirror plane (inset).}
\label{fig:2}
\end{figure*}

Fig.~\ref{fig:1}(a) demonstrates the linear polariton dispersion when the bending mode of \ch{CO2} at 655 cm$^{-1}$ is resonantly coupled to a 1D Fabry--P\'erot cavity at normal incidence. With an effective light-matter coupling strength per molecule as $\widetilde{\varepsilon}=1\times 10^{-6}$ a.u., the resulting normal-incident Rabi splitting becomes 74 cm$^{-1}$. Here, 720 liquid \ch{CO2} molecular dynamics (MD) simulation cells are evenly placed along the cavity mirror plane of length $L_x = 5$ mm. The cavity is modeled using 720 normal modes with equidistant in-plane wave numbers in momentum space with $k_{\parallel} \in [1, 720]$ cm$^{-1}$. Each MD simulation cell, as propagated via the LAMMPS MD package \cite{Thompson2022}, consists of 36 \ch{CO2} molecules in liquid phase at 300 K under periodic boundary conditions. See SI Sec. S2 for simulation details.

Fig. \ref{fig:1}(b) reports the nonequilibrium polariton dispersion when the lower polariton (LP) branch is excited at the in-plane wave number $k_{\parallel}^{\text{max}}=\omega_{\parallel}/c_0=720$ cm$^{-1}$ using a continuous wave (cw) pulse, where $c_0$ denotes the speed of light in vacuum.  
This cw pulse, with an amplitude $E_0=3.2\times 10^{-2}$ a.u., is spatially confined within a length of 1.5 mm at the center of the cavity mirror plane, corresponding to 30\% of the simulated cavity plane area. Clearly, as shown in Fig. \ref{fig:1}(b), most of the polariton population is centered at the blue-shifted LP band edge near $k_{\parallel}\approx 0$. This blue-shifted band-edge population accumulation, together with the pumping threshold and spectral narrowing shown below (Fig. \ref{fig:2}), constitutes the standard experimental signature of polariton condensation  \cite{kasprzakBoseEinsteinCondensation2006,utsunomiyaObservationBogoliubovExcitations2008,amoCollectiveFluidDynamics2009}. The corresponding photonic energy dynamics in  Fig.~\ref{fig:1}(c) reveal overwhelming energy accumulation on the band-edge cavity photon modes (brown line), also supporting the observation of LP condensation. 

In this calculation,  we have applied a Langevin thermostat with a lifetime of $\tau=5$ ps to both the molecular and photonic degrees of freedom. This cavity lifetime is consistent with typical cavity losses ($1\sim 10$ ps) for vibrational polaritons in IR Fabry--P\'erot cavities \cite{Xiang2018,Xiang2020Science}, whereas the vibrational dissipation lifetime is introduced to account for the energy relaxation to other material degrees of freedom not included in our small-scale molecular simulations. The balance between the cw pulse driving and cavity as well as molecular dissipation facilitates the formation of the $k_{\parallel}\sim 0$ LP steady state [Fig. \ref{fig:1}(b)].

By contrast, for the polariton formation using the asymmetric stretching mode at 2320 cm$^{-1}$ [Fig. \ref{fig:1}(d)], as shown in Fig. \ref{fig:1}(e), strong cw pulse pumping at $k_{\parallel}^{\text{max}}=2520$ cm$^{-1}$ does not yield selective population at the $k_{\parallel}\approx 0$ LP band edge. Compared to the equilibrium polariton dispersion [Fig. \ref{fig:1}(d)], both the LP and upper polariton (UP) branches become significantly broadened due to pump-induced in-plane translational symmetry breaking \cite{Li2024Symmetry}. The corresponding photonic energy dynamics in Fig. \ref{fig:1}(f) also suggest significantly weaker photonic energy gain at small $k_{\parallel}$ values.

The distinction between bending and stretching modes in Fig. \ref{fig:1} highlights the first constraint of vibrational polariton condensation: \textit{LP formed by the anharmonic stretching mode may not be an optimal candidate for polariton condensation}. This is because, for sustaining a steady-state macroscopic population of the $k_{\parallel}\approx 0$ LP mode, this LP state should not possess strong material dissipation pathways that dominate over the cavity loss channel. For anharmonic stretching modes forming polaritons, however, it is well known that the corresponding LP has a nonlinear absorption pathway for efficiently converting two LP quanta to localize  $0\rightarrow2$ vibrational excitations due to molecular anharmonicity \cite{Li2020Nonlinear,Ribeiro2020,Xiang2019}. Consequently, this LP-enhanced molecular nonlinear absorption pathway becomes more significant once the LP population is large, thus directly competing with and likely destroying the LP condensates. As molecular bending modes are mostly harmonic, the absence of LP-enhanced molecular nonlinear absorption facilitates the formation of LP condensation.

For gaining further insights into the mechanism of vibrational polariton condensation, Fig.~\ref{fig:2}(a) presents the polariton intensity of the LP band edge with respect to the pumping power $|E_0|^2$ on a logarithmic scale. We compare the bending modes of four  molecular liquids forming polaritons, including liquid \ch{CO_2} (blue), \ch{CS_2} (red), \ch{H_2O} (green) and \ch{CH_4} (magenta), with  the associated linear polariton spectra in SI Fig. S2. Taking the \ch{CO_2} bending mode as an example, the LP band-edge intensity increases gradually at relatively weak pumping ($E_0 \lesssim1\times10^{-2}$ a.u.). As the pumping amplitude further increases, the LP band-edge intensity rises sharply. This threshold-like behavior is another characteristic signature of nonequilibrium polariton condensation \cite{kasprzakBoseEinsteinCondensation2006}, i.e., beyond a pumping strength threshold, the redistribution of the polariton population to the band edge can no longer be described as a linear-response process; instead, the rate constant of polariton-polariton scattering from high $k$-vector states toward the LP band edge becomes significantly enhanced.  

We next compare different molecules to examine how polariton condensation depends on molecular properties, a question which is difficult to answer when only phenomenological parameters are introduced for describing the molecules. As shown in Fig. \ref{fig:2}(a), liquid \ch{CS_2} at 300 K, which has a molecular structure similar to \ch{CO_2}, exhibits a condensation threshold at a remarkably smaller field amplitude ($E_0\sim 6\times 10^{-3}$ a.u.) than that of \ch{CO2}. Beyond this threshold, it also shows a qualitatively similar, superlinear LP condensation trend with respect to the field amplitude when $E_0 \lesssim 1.6 \times 10^{-2}$ a.u. Of course, when the amplitude further grows ($E_0 \gtrsim 1.6\times10^{-2}$ a.u.), due to the thermalization of the molecular system (\ch{CS2} bending frequency is 435 cm$^{-1}$), the LP condensation saturates and can no longer grow superlinearly with respect to the pulse amplitude. As \ch{CS_2}  stays in the liquid phase under ambient conditions, this result suggests that it may serve as a better candidate for LP condensation than \ch{CO2}, which requires high pressure to stay in the liquid phase at room temperature. By contrast, for liquid \ch{H_2O} and \ch{CH_4}, Fig. \ref{fig:2}(a) also highlights that analogous condensation behaviors do not occur, suggesting their limited capacities for condensation. 

Why do these bending modes show such a fundamental difference in forming the LP condensates? This difference may relate to the rate of intramolecular vibrational energy redistribution of the excited bending modes. On the one hand, for \ch{CS_2}, the Fermi resonance between twice the bending mode and the IR-inactive symmetric stretching mode is much less likely to occur than that for \ch{CO2} due to their different resonance conditions \cite{fermiUeberRamaneffektKohlendioxyds1931,nielsenVibrationRotationEnergiesMolecules1951}. This is consistent with the reduced condensation threshold for \ch{CS2} with respect to \ch{CO2}. On the other hand, previous calculations suggest that the liquid \ch{CH4} exhibits strong and abundant vibrational relaxation channels upon the polariton excitation \cite{jiSelectiveExcitationIRInactive2025a}. For liquid \ch{H2O}, due to its strong anharmonic hydrogen bonding network (or inhomogeneously broadened vibrational spectra), the intramolecular vibrational energy redistribution typically occurs on the timescale of $\sim 10^2$ fs \cite{Nagata2015}, whose rate significantly exceeds the cavity loss rate and likely destroys any excess LP population. Overall, Fig. \ref{fig:2}(a) indicates a second constraint of vibrational polariton condensation: \textit{The molecular bending mode involved in the LP condensation should not exhibit strong intramolecular vibrational relaxation.} In other words, molecules exhibiting strong bending-to-stretching Fermi resonance or inhomogeneous broadening may not optimally form the LP condensation.

The nonequilibrium lineshapes of polariton condensation warrant careful examination. For example, during the cw pulse amplitude sweep of \ch{CS2} bending mode forming strong coupling, the nonequilibrium LP lineshape of the lowest, $k_{\parallel}\sim 0$ mode blueshifts [Fig.~\ref{fig:2}(b)], in agreement with the observation in Fig. \ref{fig:1}(b). Moreover, the LP lineshape becomes narrower under larger amplitudes, confirming again the formation of LP condensation.

\begin{figure*}[h]
\centering
\includegraphics[width=0.8\linewidth]{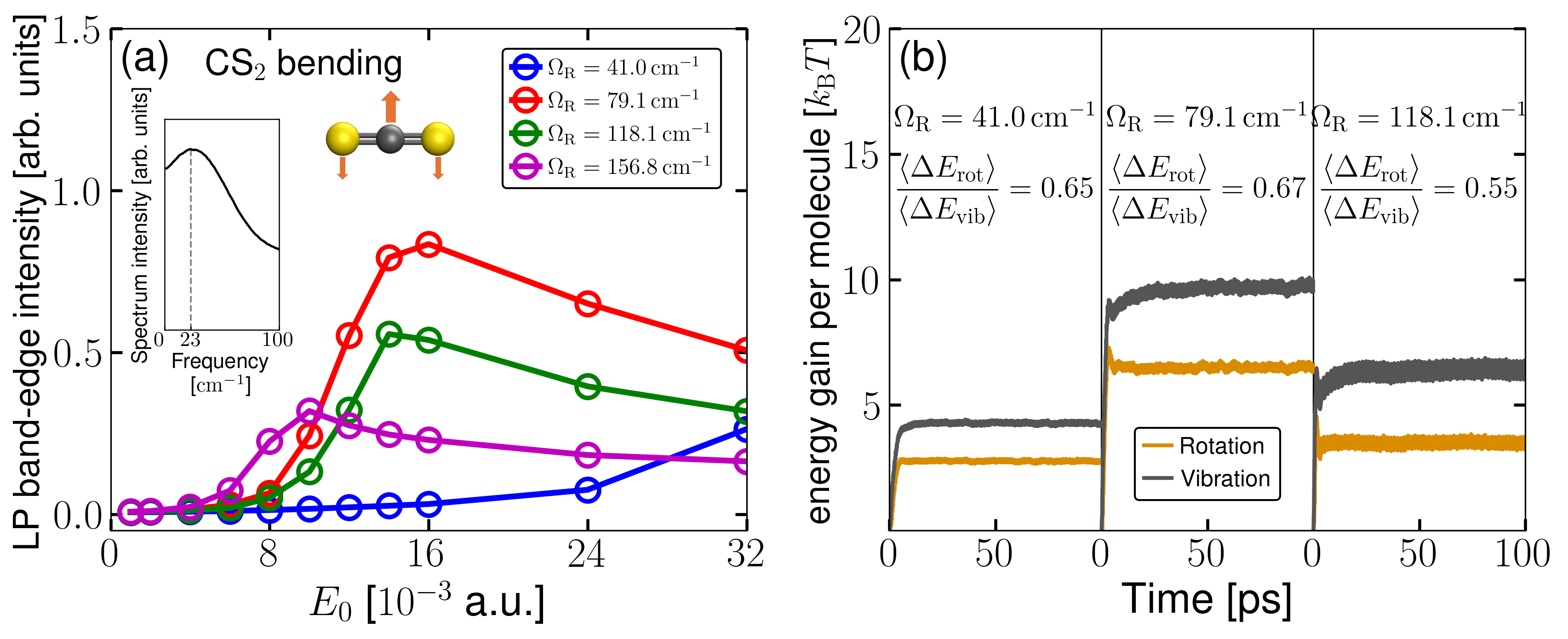}
\caption{Rabi splitting dependence of LP condensation for  the \ch{CS2} bending mode forming polaritons. (a) Band-edge LP intensity versus the pumping amplitude $E_0$ under varied Rabi splittings with $\Omega_{\rm R}=$ 41 (blue), 79 (red), 118 (green), and 157 (magenta) cm$^{-1}$. (b) Energy gain per molecule during the LP condensation formation for the vibrational (black) and rotational (brown) degrees of freedom under three sets of Rabi splittings: (from left to right) $\Omega_{\rm R}=$ 41, 79, and 118 cm$^{-1}$.  
}
\label{fig:rabi_splitting_dependence}
\end{figure*}

Apart from changing the molecules for forming strong coupling, another widely used strategy for facilitating the condensation formation is to create spatial trapping potentials, which has been discussed extensively in the exciton-polariton condensation literature \cite{baliliBoseEinsteinCondensationMicrocavity2007,askitopoulosPolaritonCondensationOptically2013,wurdackEnhancingGroundStatePopulation2022,weiOpticallyTrappedRoom2022}. 
This strategy may increase the local polariton density and reduce the condensation threshold. Inspired by this idea, we further examine the vibrational polariton thresholds for liquid \ch{CO2} and \ch{CS2}  systems when their molecular density distributions along the cavity mirror plane follow a non-uniform Gaussian distribution, as shown in the inset of Fig. \ref{fig:2}(c). This in-plane density inhomogeneity may be prepared by spin-coating an  uneven \ch{SiO2} film on the cavity mirrors, so the molecular liquid injected in the cavity naturally exhibits an inhomogeneous in-plane density distribution.

Fig.~\ref{fig:2}(c) provides the LP band-edge intensity scan over the pumping power $|E_0|^2$ for the liquid \ch{CO2} and \ch{CS2} molecular systems with broken in-plane translational symmetry. 
In line with recent work \cite{jiNonlinearFreezingVibrational2026a}, this symmetry breaking results in the enhanced polariton-polariton scattering on the LP branch from high $k$-vector states towards band-edge states. This enhanced polariton scattering greatly reduces the pumping threshold required for condensation as compared to Fig. \ref{fig:2}(a). For liquid \ch{CS_2}, the required pumping power can be reduced by up to one order of magnitude with this in-plane translational symmetry breaking. Nevertheless, for achieving vibrational polariton condensation, the required local excitation of molecules remains substantial, reaching, e.g., $2\times 10^3$ cm$^{-1}$ excess vibrational energy per molecule for \ch{CS2}  (see SI Fig. S3  for details). Maintaining molecular systems under such strong driving conditions therefore remains an important challenge for future experiments, including the preparation of cavity setups which can sustain hot and dense molecular gases due to the strong driving.

\begin{figure*}[h]
\centering
\includegraphics[width=0.8\linewidth]{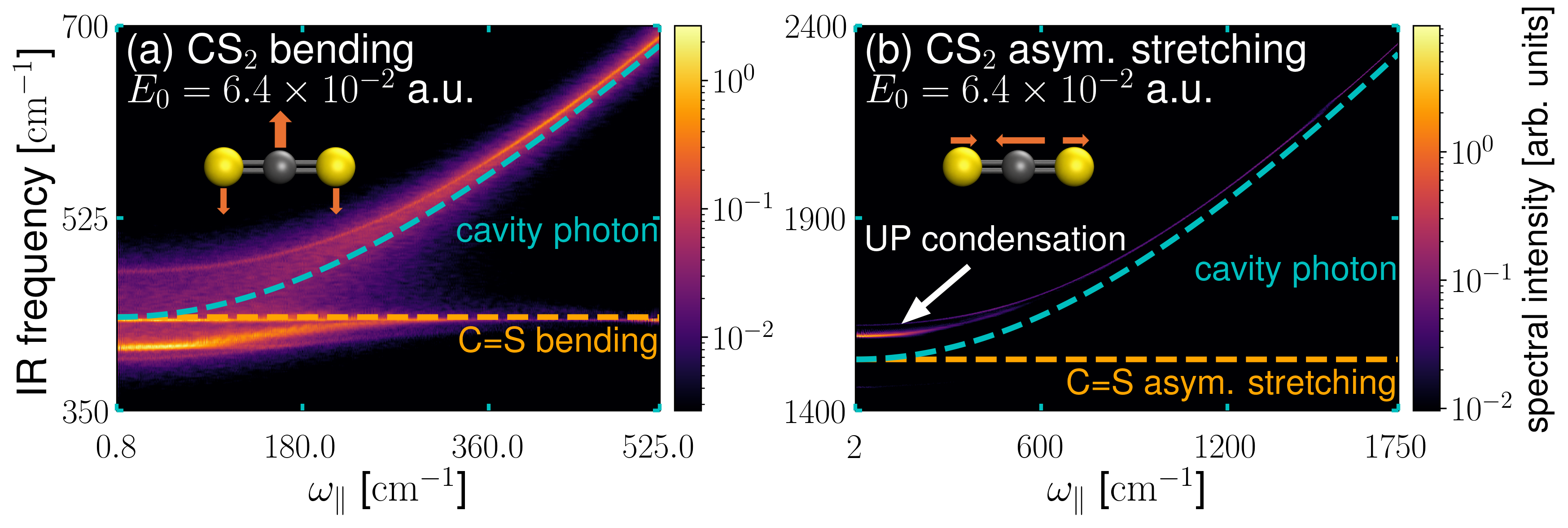}
\caption{Simulated nonequilibrium polariton dispersion for liquid \ch{CS_2} under extremely strong pumping ($E_0 = 6.4\times 10^{-2}$ a.u.) when the (a)  bending mode or (b) asymmetric stretching mode forms polaritons.  }
\label{fig:3}
\end{figure*}

While the above study highlights that harmonic, homogeneously broadened molecular bending modes may serve as good candidates for vibrational polariton condensation, one additional key observation in Fig. \ref{fig:1}(c) is that the band-edge LP modes seem to directly gain energy from  LPs at large $k_{\parallel}$ values driven by the cw pulse excitation, with the intermediate LP modes being rarely excited. This result is reminiscent of the organic exciton-polariton lasing experiments by K\'ena-Cohen and Forrest \cite{kena-cohenRoomtemperaturePolaritonLasing2010}. In their work, the polariton lasing near the LP $k_{\parallel}\sim 0$ band edge becomes directly strengthened once the frequency gap between exciton excitation and the LP band edge matches the vibrational frequency which couples to the exciton transition. Then a natural question to ask is: for vibrational polariton condensation, will an analogous mechanism also play an important role here?

For answering this question, Fig. \ref{fig:rabi_splitting_dependence}(a) plots the Rabi splitting dependence of \ch{CS2} bending-mode condensation under varied cw pulse amplitudes $E_0$, with the associated linear spectra provided in SI Fig. S6. This result suggests that 
overall the optimal condensation occurs at an intermediate Rabi splitting  when $\Omega_{\rm R} = 79$ cm$^{-1}$ (red line), a result which can also be observed in the \ch{CO2} bending-mode condensation (SI Fig. S7).
As previous work suggests that the librational (low-frequency frustrated rotations) band of liquid \ch{CS2} peaks around 30 cm$^{-1}$ \cite{Lynden-Bell2016} [which agrees with our simulated broad rotational  band that peaks at 23 cm$^{-1}$, Fig. \ref{fig:rabi_splitting_dependence}(a) inset], the libration-bending energy transfer may play a role in facilitating the LP condensation. In other words, once the frequency gap between the large-$k_{\parallel}$ driven LPs ($\sim$430 cm$^{-1}$) and the blue-shifted LP band edge [$397\sim407$ cm$^{-1}$, Fig. \ref{fig:2}(b)] matches the peak frequency of the  \ch{CS2}  librational mode, the bending-mode condensation at the band edge becomes more likely to happen.

This postulated libration-bending energy transfer mechanism is further supported by Fig. \ref{fig:rabi_splitting_dependence}(b), in which the energy gain dynamics of the rotational (brown) and vibrational (black) modes of the \ch{CS2} molecules are analyzed during the LP condensation under varied Rabi splittings when the cw pulse amplitude is fixed to $E_0=1.2\times 10^{-2}$ a.u. At $\Omega_{\rm R} = 79$ cm$^{-1}$ (middle panel), the rotational degrees of freedom receive an excess of energy relative to the vibrational energy gain, while reducing or increasing the Rabi splitting leads to weakened rotational excitation. This energy analysis result suggests a \textit{libration-bending-energy-transfer facilitated LP condensation}.

The saturation limit of the \ch{CS_2} LP condensation [Fig. \ref{fig:2}(a)] deserves further exploration. At the deep saturation limit ($E_0 = 6.4\times 10^{-2}$ a.u.), the corresponding nonequilibrium polariton dispersion in Fig. \ref{fig:3}(a) highlights the emergence of the UP  population.
When the system is driven far beyond the condensation threshold, the population redistribution is no longer confined primarily to the LP branch. Instead, higher-energy UP states become increasingly accessible due to thermalization, as the nonequilibrium temperature of molecules exceeds 1500 K here (SI Fig. S3).
Very interestingly, at the extremely strong cw pulse excitation of the $k_{\parallel}=1526$ cm$^{-1}$ LP mode formed by the \ch{CS2} asymmetric stretching mode, as shown in Fig. \ref{fig:3}(b), only the UP condensation is observed, consistent with the previous finding of nonlinear freezing of vibrational polariton transport \cite{jiNonlinearFreezingVibrational2026a}. The lack of the LP condensation here, as explained in the discussion of Fig. \ref{fig:1}, is due to the LP-enhanced molecular nonlinear absorption, which tends to destroy the high LP population formed by the anharmonic stretching mode.
In both cases, the suppression of LP condensation and  the emergence of UP band-edge population at extremely strong pulse amplitudes resemble exciton-polariton condensation experiments on switching between LP and UP condensation  by changing temperature \cite{alnatahControlledSwitchingBoseEinstein2026a}. 
This result suggests that the mesoscale CavMD approach can naturally capture polariton condensation on either polariton branch, in contrast to the Gross-Pitaevskii picture, which includes only the LP branch in the model.

\begin{figure*}
    \centering
    \includegraphics[width=\linewidth]{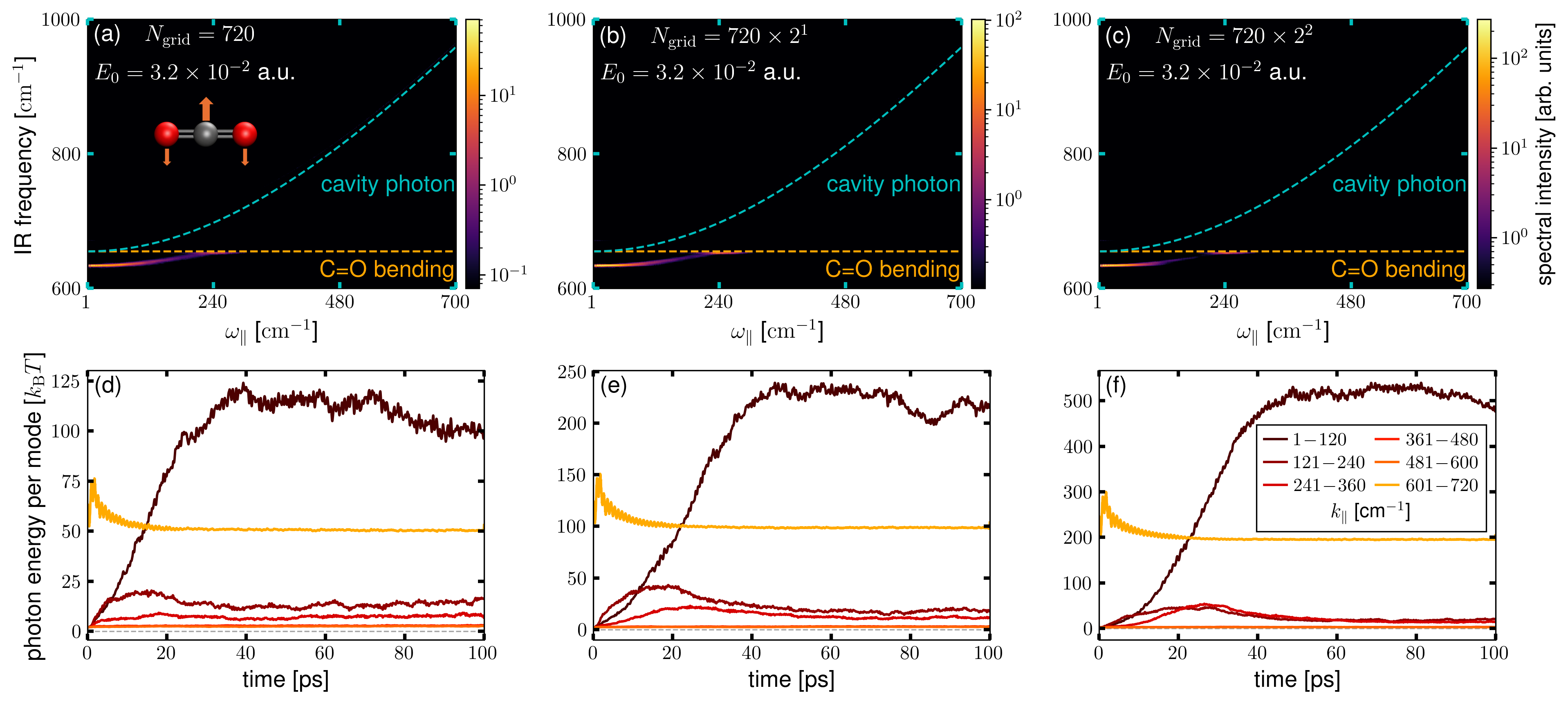}
    \caption{System size dependence of nonequilibrium polariton dispersion (top panel) and photonic energy dynamics (bottom panel) simulated with different numbers of \ch{CO_2} molecules. From left to right, the number of molecular grid points increases as $N_{\rm grid}=720\times2^n, \ (n=0,1,2)$, respectively, while the light-matter coupling strength reduces proportionally to $1/\sqrt{N_{\rm {grid}}}$ for maintaining the same Rabi splitting. The LP condensation persists with larger molecular systems.}
    \label{fig:n-dependence}
\end{figure*}

Admittedly, the mesoscale CavMD approach falls short of describing the macroscopic molecular number in optical cavity experiments. However, the reported LP condensation result persists when further increasing the molecular system size, as shown in Fig. \ref{fig:n-dependence}. For the enlarged molecular systems, the light-matter coupling per molecule is reduced proportionally to $1/\sqrt{N_{\rm{simu}}}$, where $N_{\rm{simu}}$ denotes the explicitly simulated molecular number in the CavMD propagation, so the Rabi splitting remains the same value. Moreover, as the system size increases, the energy gain per molecule is fixed in our simulations. This persistent molecular energy gain maintains the radiative pumping of individual molecules to the LP band edge, a mechanism highlighted by P\'erez-S\'anchez and Yuen-Zhou  as the driving force of exciton-polariton condensation for organic molecules \cite{perez-sanchezRadiativePumpingVs2025}. The persistence of LP condensation suggests that our mesoscale results carry over to the macroscopic limit.

\begin{figure}
\centering
\includegraphics[width=1.0\linewidth]{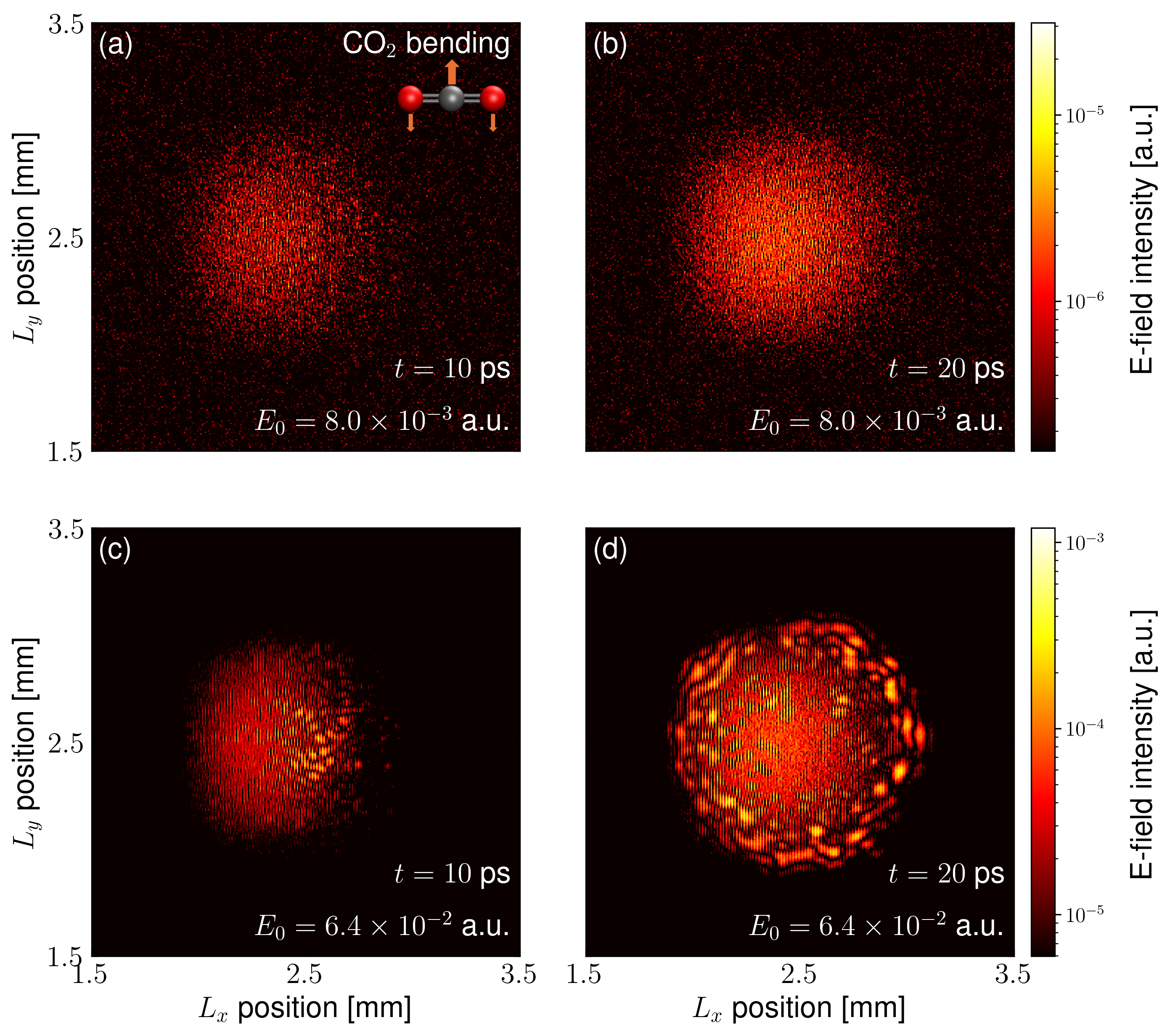}
\caption{Simulated 2D real-space imaging of $E$-field intensity distribution for the bending mode of \ch{CO2} forming polaritons. (a,b) Snapshots of $E$-field intensity distribution under weak pumping  at (a) 10 and (b) 20 ps, respectively. (c,d) Corresponding E-field intensity distribution under  strong pumping when the LP condensation forms. For parameters, a cavity setup containing a grid of $720\times720$ points, each representing an independent GPU MD engine with 36 explicitly simulated \ch{CO_2} molecules, couples to a multimode cavity along the $xy$-plane.  }
\label{fig:4}
\end{figure}

Finally, we investigate the real-space 2D imaging of vibrational polariton condensation by performing 2D mesoscale CavMD simulations of liquid \ch{CO2} bending mode under strong coupling. Here, a total of \textbf{$720\times 720$} molecular grid points are coupled to the same number of cavity modes, amounting to the explicit simulation of 55,987,200 atoms. Such a large grid is needed for a proper spatial resolution; see the SI Fig. S4 and Movie S2 for the corresponding snapshots with a \textbf{$144\times 144$} molecular grid.
Similarly to the 1D case, an analogous spatially localized cw pulse is applied to the center of the cavity mirror plane at the in-plane cavity frequency $k_{\parallel}=652$ cm$^{-1}$ along the $x$-axis.
Different from previous results, which use LAMMPS to propagate the molecules  \cite{Li2024CavMD,Ji2026MaxwellLink,jiNonlinearFreezingVibrational2026a}, we implement the GPU-accelerated molecular drivers within the \MaxwellLink package, thus enabling large-scale calculations beyond the previous capacity of CavMD. The consistency between LAMMPS and our in-house GPU code is examined in SI Fig. S5. 

Fig.~\ref{fig:4} displays two snapshots of the E-field intensity distribution over the 2D cavity mirror plane for the bending mode of \ch{CO2} forming polaritons. Here, each pixel represents cavity modes coupled to a realistic MD simulation cell.  At the weak pumping limit  ($E_0=8\times 10^{-3}$ a.u., top panel), 
the spatial distribution of the $E$-field intensity remains a Gaussian hot spot and oscillates periodically following the external pulse [Figs.~\ref{fig:4}(a,b)]. Under a strong pulse amplitude $E_0=6.4\times 10^{-2}$ a.u., instead, the $E$-field distribution starts to exhibit some fluid-like spatial features centered at the pumping hot spot [$t=10$ ps, Fig.~\ref{fig:4}(c)]. Then, these fluid-like patterns propagate to the boundary and persist throughout the whole simulation, as shown in Fig.~\ref{fig:4}(d). The associated movies of Fig. \ref{fig:4} are also available in  SI Movie S1.

One possible origin of these fluid-like patterns is the delayed excitation of cavity photon modes at intermediate $k_{\parallel}$ values following the off-resonant cw excitation at $k_{\parallel} = 652$ cm$^{-1}$, whose spatial Fourier transform exhibits these interference patterns. Indeed, these intermediate cavity modes may gain weak transient excitations during the population transfer  to the $k_{\parallel}\sim 0$ LP band edge, as indicated in Fig. \ref{fig:1}(c).
Despite the well-known concept of polariton condensates  as quantum fluids of light, this classical simulation suggests that some spatial fluid-like features of the E-field distribution could be attributed to classical effects  \cite{amoCollectiveFluidDynamics2009,amoPolaritonSuperfluidsReveal2011,lerarioRoomtemperatureSuperfluidityPolariton2017}. This is consistent with the generalized Gross-Pitaevskii equation modeling of polariton condensation \cite{woutersExcitationsNonequilibriumBoseEinstein2007}, in which a classical or mean-field field theory produces these effects naturally.

\section*{Discussion}

To summarize, we have provided key design insights of vibrational polariton condensation by fully atomistic simulations of the extended mesoscale light-matter system in Fabry--P\'erot cavities. The LP condensation may exist for harmonic bending modes of simple, homogeneously broadened liquid systems forming strong coupling. We have also quantified a possible route to lower the pumping threshold of condensation by enforcing in-plane symmetry breaking. The important role of libration-bending coupling in facilitating the LP condensation has also been examined.

For the chemical physics audience, forming a polariton condensate provides a new means for preparing a nonequilibrium vibrationally excited-state distribution in liquids beyond heating, which sheds light on pulse-controlled liquid-phase chemistry.
From a methodology perspective, this work establishes mesoscale CavMD simulations as a novel predictive approach for understanding polariton condensation. The unique strength of this atomistic approach is its predictive capacity for revealing how intrinsic molecular properties influence the condensation formation beyond phenomenological models. 
Together with previous work, mesoscale CavMD offers a unified framework for understanding vibrational polariton dynamics in planar Fabry--P\'erot cavities, ranging from polariton relaxation to dark modes, the ballistic-to-diffusive turnover and the nonlinear freezing of polariton transport \cite{jiNonlinearFreezingVibrational2026a}, as well as the necessary conditions for forming nonequilibrium polariton condensation. 
Its integration with nonadiabatic molecular dynamics may further enable the direct atomistic understanding of exciton-polariton condensation, an exciting direction to explore in the future.

\section{Data Availability Statements}

Code and processed datasets have been deposited in Zenodo and will become publicly available upon the publication of this work.

\section{Acknowledgments}

This material is based upon work supported by the U.S. National Science Foundation under Grant No. CHE-2502758. This work used the Anvil HPC at Purdue University through allocation CHE250091 from the Advanced Cyberinfrastructure Coordination Ecosystem: Services \& Support (ACCESS) program \cite{Boerner2023}, which is supported by U.S. National Science Foundation grants \#2138259, \#2138286, \#2138307, \#2137603, and \#2138296.


\providecommand{\latin}[1]{#1}
\makeatletter
\providecommand{\doi}
  {\begingroup\let\do\@makeother\dospecials
  \catcode`\{=1 \catcode`\}=2 \doi@aux}
\providecommand{\doi@aux}[1]{\endgroup\texttt{#1}}
\makeatother
\providecommand*\mcitethebibliography{\thebibliography}
\csname @ifundefined\endcsname{endmcitethebibliography}  {\let\endmcitethebibliography\endthebibliography}{}

\end{document}